%% file: main.tex
\documentclass[10pt,conference]{IEEEtran}

\usepackage{algorithm}
\usepackage{algpseudocode}
\usepackage{makeidx}  
\usepackage[dvips]{graphicx}
\usepackage{booktabs}
\usepackage{url}
\usepackage{amssymb}
\usepackage{amsmath}
\usepackage{color}
\usepackage{epstopdf}
\newtheorem{thm}{Theorem}

\newcommand{\BigO}[1]{\ensuremath{\mathcal{O}(#1)}}

\begin{document}

\title{EMinRET: Heuristic for Energy-Aware VM Placement with Fixed Intervals and Non-preemption}

\author{\IEEEauthorblockN{Nguyen Quang-Hung,
 Nam Thoai}\\
\IEEEauthorblockA{Faculty of Computer Science and Engineering, \\
HCMC University of Technology, VNU-HCM \\
Ho Chi Minh City, Vietnam\\
Email: \{hungnq2,nam\}@cse.hcmut.edu.vn}
}

\maketitle              

\input{abstract}


\section{Introduction}
\label{sec:intro}
\input{intro} 
\section{Related Works}
\label{sec:related}
\input{related}
\section{Problem Description}
\label{sec:problem}
\input{problem}
\input{theorems}
\section{EMinRET: Energy-Aware Minimizing Resource Efficiency - Time Heuristic}
\label{sec:algoEMinRET}
\input{algo}
\section{Performance Evaluation}
\label{sec:experiment}
\input{experiment}
\section{Conclusions and Future Work}
\label{sec:concl}
\input{concl}

\section*{Acknowledgment}
This research was conducted within the "Studying and developing practical heuristics for energy-aware virtual machine-based lease scheduling problems in cloud virtualized data centers" sponsored by TIS.
\bibliographystyle{IEEEtran} 
\bibliography{refs}



\end{document}

%% file: abstract.tex
\begin{abstract}
Infrastructure-as-a-Service (IaaS) clouds have become more popular enabling users to run applications under virtual machines.
This paper investigates the energy-aware 
virtual machine (VM) allocation problems in IaaS clouds along characteristics:
multiple resources, 
and fixed interval times and non-preemption of virtual machines.
Many previous works proposed to use a minimum number of physical machines;
however, this
is not necessarily a good solution to minimize total energy consumption in the VM placement with multiple resources, fixed interval times and non-preemption.
We observed that minimizing total energy consumption of physical machines
is equivalent to minimize the sum of total completion time of all
physical machines.
Based on the observation,
we propose EMinRET algorithm.
The EMinRET algorithm swaps an allocating VM with a suitable overlapped VM,
which is of the same VM type and is allocated
on the same physical machine,
to minimize total completion time of all physical machines.
 The EMinRET uses resource utilization during executing time period of a physical machine
as the evaluation metric,
 and will then choose a host that minimizes the metric to allocate a new VM.
 In addition, this work studies some heuristics for sorting the list of virtual machines 
 (e.g., sorting by the earliest starting time, or the longest duration time first, etc.)
to allocate VM.
Using the realistic log-trace in the Feitelson's Parallel Workloads Archive,
our simulation results show that the EMinRET algorithm could reduce from 25\% to 45\%
  energy consumption compared with
  power-aware best-fit decreasing (PABFD) \cite{Beloglazov2011}) and
  vector bin-packing norm-based greedy algorithms (VBP-Norm-L1/L2 \cite{Panigrahy2011}).
  Moreover, the EMinRET heuristic has also less total energy consumption than our previous heuristics (e.g. MinDFT and EPOBF) in the simulations 
   (using same virtual machines sorting method).

\end{abstract}

%% file: intro.tex
IaaS cloud systems are often built from virtualized data centers
\cite{Sotomayor2010,Beloglazov2012,barroso2013datacenter}.
Power consumption in a large-scale data centers requires multiple megawatts \cite{Fan2007a,Le2011}. 
Le et al. \cite{Le2011} estimate the energy cost of a single data center is more than \$15M per year. 
As these data centers scale, they will consume more energy.
Therefore, advanced scheduling techniques for
reducing energy consumption of these cloud systems are highly
concerned for any cloud providers to reduce energy cost. 
Increasing energy cost and the need to environmental sustainability address energy efficiency is a hot research topic in cloud systems.
Energy-aware scheduling of VMs in IaaS cloud is still challenging \cite{Garg2009,Le2011,Quang-Hung2014,Tako2012,Vis2011}.

Many previous works \cite{Beloglazov2012,Beloglazov2011,Panigrahy2011} proved that the virtual machine allocation is NP-Hard
and proposed to address the problem of 
energy-efficient scheduling of VMs in cloud data centers. 
They \cite{Beloglazov2012,Beloglazov2011,Panigrahy2011} present
techniques for 
consolidating virtual machines in cloud data centers by using 
bin-packing heuristics (such as First-Fit Decreasing \cite{Panigrahy2011}, and/or Best-Fit Decreasing \cite{Beloglazov2011}).
They attempt to minimize the number of running physical machines
and to turn off as many idle physical machines as possible. 
Consider a d-dimensional resource allocation
where each user requests a set of virtual machines (VMs). 
Each VM requires multiple resources (such as CPU, memory, and IO)
and a fixed quantity of each resource at a certain time interval.
Under this scenario, using a minimum of physical machines may not be a good solution.
Our observations are that
using a minimum number of physical machines
is not necessarily a good solution to minimize total energy consumption.
In a homogeneous environment where all physical servers are identical,
the power consumption of each physical server is linear to its CPU utilization,
i.e.,
a schedule with longer working time will consume more energy
than another schedule with shorter working time.

\input{example}


The EMinRET heuristic places VMs that request multiple resources in the fixed interval time and non-preemption into physical machines to minimize total energy consumption of physical machines while meeting all resource requirements.
Using numerical simulations, we compare the EMinRET 
with the popular modified best-fit decreasing (PABFD) \cite{Beloglazov2011},
two vector bin-packing norm-based greedy (VBP-Norm-L1/L2) \cite{Panigrahy2011}, 
and our previous algorithms (e.g. EPOBF-ST/FT \cite{Quang-Hung2014}, and MinDFT-ST/FT \cite{HungFDSE2014}).
Using two real log-traces (i.e., \cite{SDSCBLUEWorkload} and \cite{HPC2NWorkload}) in the Feitelson's Parallel Workloads Archive,
  our simulation results show that the EMinRET heuristic with its configurations could reduce from 25\% to 45\%
  total energy consumption compared with
  Power-Aware Best-Fit Decreasing (PABFD) \cite{Beloglazov2011} and
  two norm-based greedy heuristics (VBP-Norm-L1/L2) \cite{Panigrahy2011}.
Moreover, the EMinRET heuristic has also less total energy consumption than our previous heuristics (e.g. MinDFT and EPOBF) in the simulations 
 (using same virtual machines sorting method). 


The rest of this paper is structured as follows. 
Section \ref{sec:related} discusses related works.
Section \ref{sec:problem} describes the energy-aware VM allocation problem with multiple requested resources, fixed starting and duration time. 
We also formulate the objective of scheduling, and present our theorems. 
 The proposed EMinRET algorithm present in Section \ref{sec:algoEMinRET}.
 Section \ref{sec:experiment} discusses our performance evaluation using simulations.
 Section \ref{sec:concl} concludes this paper and introduces future works.

%% file: example.tex
\begin{table}[!hbt]
\caption{Example showing that using a minimum number of physical servers is not optimal. (*: normalized demand resources to physical server’s capacity resources)}
\label{table:example1}
\begin{center}
\begin{tabular}{|l|c|c|c|c|c|}
\hline
VM ID & CPU* & RAM* & Network* & Starttime & Dur. (hour) \\
\hline
VM1             & 0.5                             & 0.1                      & 0.2                         & 0                              & 10                                  \\
VM2             & 0.5                             & 0.5                      & 0.2                         & 0                              & 2                                  \\
VM3             & 0.2                             & 0.4                      & 0.2                         & 0                              & 2                                   \\
VM4             & 0.2                             & 0.4                      & 0.2                         & 0                              & 2                                   \\
VM5             & 0.1                             & 0.1                      & 0.1                         & 0                              & 2                                  \\
VM6             & 0.5                             & 0.5                      & 0.2                         & 1                              & 9                               \\ \hline
\end{tabular}
\end{center}
\end{table}

To the best of our knowledge, our work is the first work that studies increasing time and resource efficiency-based approach to allocate VMs onto physical machines in other that it minimizes total energy consumption of all physical machines. 
 Each VM requests resource allocation in a fixed starting time and non-preemption for the duration time.
 We present here an example to demonstrate our ideas to 
minimize total energy consumption of all physical machines in the VM placement with fixed starting time and duration time.
For example, given six virtual machines (VMs) with their resource
demands described in Table~\ref{table:example1}. 
Our constraints in the example are the maximum capacity of each resource is 1.
In the example, a bin-packing-based algorithm could result in a schedule $S_{1}$
in which two physical servers are used:
one for allocating VM1, VM3, VM4, and VM5; and another one for
allocating VM2 and VM6. The resulted total completion time
is (10 + 10) = 20 hours. However, in another schedule $S_{2}$ in which  
where VMs are placed on three physical servers,
VM1 and VM6 on the first physical server,
VM3, VM4 and VM5 on the second physical server,
and VM2 on the third physical server,
then the total completion time of the five VMs is only (10 + 2 + 2) = 14 hours.

%% file: related.tex

Many previous research \cite{Beloglazov2011,Beloglazov2012,Knauth2012,Tako2012,Chen2014} proposed algorithms that consolidate VMs 
 onto a small set of physical machines (PMs) in virtualized datacenters 
 to minimize energy/power consumption of PMs.
A group in Microsoft Research \cite{Panigrahy2011} has studied first-fit decreasing (FFD) based heuristics for vector bin-packing to minimize number of physical servers in the VM allocation problem.
 Some other works also proposed meta-heuristic algorithms to minimize the number of physical machines.
Beloglazov et al. 
 \cite{Beloglazov2011,Beloglazov2012} have proposed VM allocation problem as bin-packing problem and presented a power-aware best-fit decreasing
(denoted as PABFD) heuristic.
PABFD sorts all VMs in a decreasing order of CPU utilization and tends to
allocate a VM to an active physical server that would take the minimum
increase of power consumption. 
Knauth et al. \cite{Knauth2012} proposed the OptSched scheduling algorithm to 
 reduce cumulative machine up-time (CMU) by 60.1\% and 16.7\% in comparison to a round-robin and First-fit.
 The OptSched uses an minimum of active servers to process a given workload.
 In a heterogeneous physical machines, the OptSched maps a VM to a first available and the most powerful machine that has enough VM's requested resources. Otherwise, the VM is allocated to a new unused machine. 
  In the VM allocation problem, however, minimizing the number of used physical machines is not equal to minimizing total of total energy consumption of all physical machines.
Previous works do not consider multiple resources, fixed starting time and non-preemptive duration time of these VMs.
Therefore,
it is unsuitable for the power-aware VM allocation
considered in this paper,
i.g. these previous solutions can not result in a minimized total energy consumption
for VM placement problem with certain interval time while still
fulfilling the quality-of-service.


Chen et al \cite{Chen2014} observed there exists VM resource utilization patterns.
 The authors presented an VM allocation algorithm to consolidate complementary VMs with spatial and temporal-awareness in physical machines.
 They introduce resource efficiency and use norm-based greedy algorithm, which is similar to in \cite{Panigrahy2011}, to measure 
 distance of each used resource's utilization and maximum capacity of the resource in a host. 
 Their VM allocation algorithm selects a host that minimizes the value of this distance metric to allocate a new VM.
 Our proposed EMinRET uses a different metric that unifies both increasing time and resource efficiency. 
 In our proposed metric, the increasing time is the difference between two completion time of a host after and before allocating a VM. 
 In addition, our proposed EMinRET core algorithm has swapping step of overlapped VMs together 
 to minimize total completion times of all physical machines.

Some other research \cite{Garg2009,Le2011,Tako2012} considered
HPC applications/jobs in HPC clouds. 
Garg et al. \cite{Garg2009} proposed a meta-scheduling problem
to distribute HPC applications to cloud systems with distributed
$N$ data centers. The objective of scheduling is minimizing $CO_{2}$
emission and maximizing the revenue of cloud providers.
Le et al. \cite{Le2011} distribute VMs across distributed
cloud virtualized data centers whose electricity prices
are different in order to reduce
the total electricity cost.
Takouna et. al., \cite{Tako2012} presented power-aware
multi-core scheduling and their
VM allocation algorithm selects a host which has the minimum
increasing power consumption to assign a new VM. 
The VM allocation algorithm, however, is similar to
the PABFD’s \cite{Beloglazov2011} except that it concerns
memory usage in a period of estimated runtime for estimating the host's energy. 
The work also presented a method to select optimal operating
frequency for a (DVFS-enabled) host and configure the number of
virtual cores for VMs. 
Our proposed EMinRET algorithm that
differs from these previous works. Our EMinRET algorithm use the VM's fixed
starting time and duration time to minimize the total working
time on physical servers, and consequently minimize the total
energy consumption in all physical servers. 
To the best of our knowledge, no existing works that surveyed in \cite{BelBuLZA2010Taxonomy,orgerie2014survey,Hameed2014,IvonaSurvey2014} have thoroughly considered these aspects in addressing the problem of VM placement.


In 2007, Kovalyov et al. \cite{kovalyov2007fixed} has presented a work to describe characteristics of a fixed interval scheduling problem in which each job has fixed starting time, fixed processing time, and is only processed in the fixed duration time on a available machine. The scheduling problem can be applied in other domains. 
Angelelli et al. \cite{Angelelli20113650} considered interval scheduling with a resource constraint in parallel identical machines. 
 The authors proved the decision problem is NP-complete if number of constraint resources in each parallel machine is a fixed number greater than two.

%% file: problem.tex
\subsection{Notations}
We use the following notations in this paper:

$ vm_{i}$: The $i^{th}$ virtual machine to be scheduled.

$ M_{j}$: The $j^{th}$ physical server.

$ S $: A feasible schedule. 

$ P_{j}^{idle}$: Idle power consumption of the $M_{j}$.

$ P_{j}^{max} $: Maximum power consumption of the $M_{j}$.

$ P_{j}(t) $: Power consumption of the ($M_{j}$) at a time point $t$.


$ ts_{i}$: Fixed starting time of $vm_{i}$.

$ dur_{i}$: Duration time of $vm_{i}$.

$ T $: Maximum schedule length, which is the time that the last virtual machine will be finished.

$ n_{j}(t) $: Set of indexes of all virtual machines that are assigned to the $M_{j}$ at time $t$.

$T_{j}$ : Total busy time (working time) of the $M_{j}$.

 $ e_{i} $: Energy consumption for running the $ vm_{i}$ in the physical machine that the $ vm_{i}$ is allocated.\\

\input{energymodel.tex}

\subsection{Problem formulation}

Given a set of virtual machines $ vm_{i}$ ($i = 1,2,...,n$) to be scheduled on a
set of physical servers $ M_{j}$ ($j = 1,2,...,m$). 
Each VM is represented
as a d-dimensional vector of demand resources, 
i.e. $vm_{i} = (x_{i,1}, x_{i,2}, ..., x_{i,d}) $.
Similarly,
each physical machine is denoted as a d-dimensional 
vector of capacity resources, 
i.e. $ M_{j} = (y_{j,1}, y_{j,2}, ...,  y_{j,d}) $.
We consider types of resources such as processing element (core), 
 computing power (Million instruction per seconds -MIPS),
 physical memory (RAM), network bandwidth (BW), and storage. 
Each $vm_{i}$ is started at a fixed starting time ($ts_{i}$) 
and is non-preemptive during its duration time ($dur_{i}$). 

We assume that the power consumption model is linear to CPU utilization.
Even if all physical servers are identical and all VMs are identical too,
the scheduling is still NP-hard with $d \ge 1$ \cite{Panigrahy2011}.
With the problem considered in this paper,
all physical servers are identical and 
their power consumption models are linear to their CPU utilization
as can be seen in the two equations (\ref{eq:power}) and (\ref{eq:energy}).
The energy consumption of a physical server in a time unit is denoted as $E_{0}$
and is the same for all physical servers since the servers are identical.
The objective is to find out a feasible schedule $S$ that minimizes
the total energy consumption
in the equation (\ref{eq:minimize}) 
 with $i \in \{1,2,...,n\}$, $j \in \{1,2,...,m\}$, $t \in [0;T]$ as following:


\begin{equation}
\label{eq:minimize}
\textbf{Minimize} \   ( E_{0} \times \sum_{j=1}^{m} T_{j}  +  \sum_{i=1}^{n} e_{i} )
\end{equation}  

{\noindent}where
the total busy time of a physical server \cite{Flammini2010}, denoted as $T_{j}$, is defined as length of union of interval times of all VMs 
 that are allocated to a physical machine $M_{j}$ at time $T$. 
 \begin{equation}
 T_{j} =  span ( \bigcup_{vm_{i} \in M_j} {[ts_{i} , ts_{i} + dur_{i}]})
 \end{equation}

{\noindent}The union of two time intervals [a,b] and [c,d] is defined as:
$[a,b] \cup [c,d] = \{ x \in \mathbb{R} ~ | ~ x \in [a,b] ~ or ~ x \in [c,d] \}$
and given a time interval $I=[a, b], a\leq b: span(I) = b-a$.
If two time interval are not overlapped then span of the two non-overlapped interval is sum of span of each interval, 
$I_1=[a,b],I_2=[c,d], a \leq b \leq c \leq d: span(I_1 \cup I_2) = span(I_1) + span(I_2) = (b-a) + (d-c)$.




The scheduling problem has the following hard constraints that are described in our previous work \cite{HungFDSE2014}.
%
%
%
%
%
%

%% file: energymodel.tex
\subsection{Power consumption model}

In this paper, we use the following energy consumption model proposed in \cite{Fan2007a} for a physical machine.
 The power consumption of the $M_{j}$, denoted as $ P_{j}(.)$, is formulated as follow:
\begin{equation}
\label{eq:power}
 P_{j}(t) = P_{j}^{idle} + (P_{j}^{max} - P_{j}^{idle}) U_{j}(t) 
 \end{equation}
 
 
 The CPU utilization of the physical server at time $t$, denoted as $ U_{j}(t)$, is defined as the average percentage of total of allocated 
 computing powers of $n_j(t)$ VMs that is allocated to the $M_j$. 
We assume that all cores in CPU are homogeneous, i.e. $\forall c=1,2,...,PE_j: MIPS_{j,c}=MIPS_{j,1}$ , The CPU utilization is formulated as follow:
 \begin{equation}
 \label{eq:cpuutilization}
 U_{j}(t) = (\dfrac{1}{PE_{j} \times MIPS_{j,1}}) \sum_{c=1}^{PE_{j}} \sum_{i  \in  n_j(t)} mips_{i,c} 
 \end{equation}

The energy consumption of the server in period time [$t_1, t_2$] is formulated as follow:
\begin{equation}
\label{eq:energy}
E_{j} = \int_{t_{1}}^{t_{2}} P_{j}( U_{j}(t)) dt
\end{equation}

{\noindent}where: \\
 $ U_{j}(t) $ : CPU utilization of the $M_{j}$ at time $ t $ and $ 0 \leq U_{j}(t) \leq 1 $. \\
 $ PE_{j}$ : Number of processing elements (i.e. cores) of the $M_{j}$.\\
 $ mips_{i,c} $	: Allocated MIPS of the c$^{th}$ processing element to the $vm_{i}$ by the $M_{j}$.\\
 $ MIPS_{j,c} $ : Maximum capacity computing power (Unit: MIPS) of the c$^{th}$ processing element on the $M_{j}$.\\


%% file: theorems.tex
\subsection{Theorems}
\label{sec:theorems}


\begin{thm}
\label{sec:theorem01}
Given a cloud system with a set of identical physical machines, 
assume that power consumption of a physical machine is $P(u) = b + au$, 
in which: 
 $b = P_{idle}$ is the idle power consumption, 
 $a = P_{max} - P_{idle}$ is the maximum power consumption, and
 $u$ is the CPU utilization in percentage ($0 \leq u \leq 1$).
 We denote $e_{ij}$ is energy consumption of each virtual machine $i$-th that is scheduled on any physical machine $j$-th.
 If the $u$ of the mapped virtual machine is a constant, 
 then the energy consumption of each virtual machine, $e_{ij}$, is independent of any mapping (i.e. any schedule). 
 We have $\forall i \in {1,2,...,n}, j \in {1,2,...,m}: e_{ij} = e_{i}$.
\end{thm}

\begin{proof}
Recall that the energy consumption is formulated in Equation (\ref{eq:energy}), 
 and power consumption, $P(u)$, is a linear function of CPU utilization, $u$. 
 Therefore for all $ i \in {1,2,...,n}$ and $j \in {1,2,...,m}$, we see that $e_{ij}$ is the integral of the $P(u)$ over any time interval [$t_{1}$, $t_{2}$],
and is the same value, denoted as $e_{i}$.
\end{proof}

From Theorem \ref{sec:theorem01}, we can imply the following theorem
to prove that Equation \ref{eq:minimize} is equivalent to Equation \ref{eq:minimize2}. 
\begin{thm}
\label{sec:theorem02}
 Minimizing total energy consumption is equivalent to
 minimizing the sum of total busy time of all physical machines ($\sum_{j=1}^{m} T_{j}$).
 
\begin{equation}
\label{eq:minimize2}
\textbf{Minimize} \   ( E_{0} \times \sum_{j=1}^{m} T_{j}  +  \sum_{i=1}^{n} e_{i} ) ~ \sim ~ \textbf{Minimize} \   ( \sum_{j=1}^{m} T_{j})
\end{equation}  

\end{thm}

\begin{proof}
According to the objective function
described in (\ref{eq:minimize}), $E_0$ is constant while
$e_i$ is independent of any mapping (i.e. any schedule). 
\end{proof}

Based on the above observations, we propose our energy-aware algorithms denoted as EMinRET
which is presented in the next section.

%% file: algo.tex
\subsection{Scheduling algorithm}
\label{sec:schedalgo}

\input{fig-algo-core}

In this section, we present our energy-aware scheduling algorithm, namely, EMinRET.
EMinRET presents a metric to unify the increasing time and estimated resource efficiency when mapping a VM onto a physical machine.
Then, EMinRET will choose a host that has minimizing of the metric. 
   Our previous MinDFT-ST/FT \cite{HungFDSE2014} only focused on minimizing the increasing time when mapping a VM onto a physical machine.
   The EMinRET additionally considers resource efficiency
   during an execution period of a physical machine in order to fully utilize resources in a physical machine.
   Furthermore, the core EMinRET algorithm can swap an overlapped VM, which has already been assigned to an active physical machine before, 
   with new VM to minimize total completion time of the physical machine. 
   In this paper, two VMs are overlapped if $ts_1 < ts_2 < (ts_1 + dur_1) < (ts_2 + dur_2)$, where $ts_1$, $ts_2$, $dur_1$, $dur_2$ are starting times and duration times of two VMs.
   The core EMinRET algorithm will swap new VM and its overlapped VM together if two VMs meet conditions: 
   	(i) both VMs are of the same VM type (i.e. the same amount of requested resources such as number of CPU core, physical memory, network bandwidth, storage, etc.); 
   	(ii) the new VM has duration time longer than its overlapped VM.
   Our previous MinDFT-ST/FT  \cite{HungFDSE2014} do not have this swapping step, and neither the EPOBF-ST/FT \cite{Quang-Hung2014} have these steps.

Based on Equation \ref{eq:cpuutilization}, the utilization of a resource $r$ (resource $r$ can be CPU, physical memory, network bandwidth, storage, etc.) of the $M_j$, denoted as $U_{j,r}$, is formulated as:
 \begin{equation}
 \label{eq:resutilization}
 U_{j,r} = \sum\limits_{s \in n_{j}} \dfrac{V_{s,r}}{ H_{j,r}}.
  \end{equation}

{\noindent}where $n_{j}$ is the list of virtual machines that are assigned to the $M_j$, 
 $V_{s,r}$ is the amount of requested resource $r$ of the virtual machine $s$ (note that in our study the value of $V_{s,r}$ is fixed for each user request), 
 and $H_{j,r}$ is maximum capacity of resource $r$ in the $M_j$.
  
Inspired by the work from Microsoft research team \cite{Panigrahy2011,Chen2014},
 resource efficiency of a physical machine $j$-th, denoted by $RE_{j}$, 
 is Norm-based distant \cite{Panigrahy2011} of two vectors: normalized resource utilization vector and unit vector $\mathbf{1}$, 
 the resource efficiency is formulated as:
\begin{equation}
\label{eq:resefficiency}
RE_{j} = \sum_{r \in \mathcal{R}} ((1 - U_{j,r}) \times w_r)^2
 \end{equation}
{ \noindent}where $\mathcal{R}$=\{cpu, ram, netbw, io, storage\}: set of resource types in a host, $w_r$ is weight of resource $r$ in a physical machine.

 In this paper, we propose a unified metric for increasing time and resource efficiency that is calculated as:
 \begin{equation}
 \label{eq:retime}
 RET = (t^{diff} \times w_{r=time})^2 + \sum_{r \in \mathcal{R}}^{} ((1 - U_{j,r}) \times w_r)^2
 \end{equation}
  
EMinRET chooses a physical host that has a minimum value of the $RET$ metric to allocate for a VM.
We present the pseudo-code of EMinRET in Algorithm~\ref{alg:eminret}.
 The EMinRET can sort the list of VMs by earliest starting time first, or earliest finishing time first, or longest duration time first, etc..
The EMinRET solves the scheduling problem in time complexity of $\BigO{n \times m \times q}$
where $n$ is the number of VMs to be scheduled, $m$ is the number of physical machines, and $q$ is the maximum number of allocated VMs in the physical machines $M_j, \forall j=1,2,...,m$.

%% file: fig-algo-core.tex
\begin{algorithm}[!ht]
\caption{EMinRET: Energy-Aware Minimizing Resource Efficiency - Time}\label{alg:eminret}
 \begin{algorithmic}[1]
\Function{EMinRET}{}
\State	\textit{Input:} vmList - a list of virtual machines to be scheduled, 
 hostList - a list of physical servers
\State	\textit{Output:} mapping (a feasible schedule) or null 
\State	vmList = sortVmListByOrder( vmList, order=[starttime, finishtime] ) \Comment{1}\label{line:sortvmlistbycriteria}
\State  m = hostList.size(); n = vmList.size();
\State T[j] = 0, $\forall j \in [1, m]$ 
\For {$i = 1$ to n} \Comment{on the VMs list}
		\State 	vm = vmList.get(i) 
		\State	allocatedHost = null 
		\State	T1 = sumTotalHostCompletionTime( T ) 
		\State	minRETime = $+\infty$
		\For {$j = 1$ to m}  \Comment{on the hosts list}
         	\State host = hostList.get(j)
         	\State hostVMList = sortVmListByOrder( host.getVms(), order=[starttime, finishtime]) 
         	\For {(Vm vmTemp : hostVMList) }
         		\If {canSwap(vm, vmTemp)} 
         				\State swap(vm, vmTemp) and break
         		\EndIf 
         	\EndFor 
			\If {host.checkAvailableResource( vm )} \\
				\State preTime = T[ host.id ]
				\State T[ host.id ] = host.estimateHostTotalCompletionTime( vm ) 
				\State T2 = sumTotalHostCompletionTime( T )
				\State RETime = EstimateMetricTimeResEff(  T2 - T1, host ) 
				\If {(minRETime $>$ RETime )}
					\State	minRETime = RETime 
					\State	allocatedHost = host 
				\EndIf 
				\State T[ host.id ] = preTime				
				\Comment{ Next iterate over hostList and choose the host that minimize the value of different time and resource efficiency } 
            \EndIf
		\EndFor 
		
		\If {(allocatedHost != null) }
			\State allocate the $vm$ to the $host$
			\State add the pair of $vm$ (key) and $host$ to the $mapping$
		\EndIf
\EndFor	
\State	return $mapping$ 
\EndFunction

\State {sumTotalHostCompletionTime}({T[]}) = $\sum_{j=1}^{m} T_{j}$ \Comment{T[1...m]: Array of total completion times of $m$ physical servers}

\end{algorithmic}
\end{algorithm}

\begin{algorithm}[ht]
\caption{Estimating the metric for increasing time and resource efficiency}\label{alg:estimretime}
 \begin{algorithmic}[1]
 \Function{EstimateMetricTimeResEff}{}
 \State	\textit{Input:} difftime - a different time,
								host - a candidate physical machine
 \State	\textit{Output:} ret - a value of metric time and resource efficiency
 \State Set $\mathcal{R}$=\{cpu, ram, netbw, io, storage, time\}
 \State $j$=host.getId(); $n_j$=host.getVMList();
 \For{$r \in \mathcal{R}$} 
 	\State Calculate the resource utilization, $U_{j,r}$ as in the Equaltion (\ref{eq:resutilization}). 
 \EndFor	
 \State $weights[] \leftarrow$ Read resource weights from configuration file.
 \State $ret = (difftime \times weights[time])^2 + \sum_{r \in \mathcal{R}}^{} ((1 - U_{j,r}) \times weights[r])^2 $  \Comment{weights[time] is weight of the different time}
 \State return $ret$
 \EndFunction 
 
\end{algorithmic}
\end{algorithm}

%% file: experiment.tex
\subsection{Algorithms}

In this section, we study the following VM allocation algorithms:

\begin{itemize}
\item
PABFD, a power-aware and modified best-fit decreasing heuristic \cite{Beloglazov2011}\cite{Beloglazov2012}.
The PABFD sorts the  list of $VM_{i}$ (i=1, 2,..., n) by
their total requested CPU utilization, and 
assigns new VM to any host that has a minimum increase in power consumption.

\item
VBP-Norm-LX, a family of vector packing heuristics that
is presented as Norm-based Greedy with degree X={1, 2} \cite{Panigrahy2011}.
Weights of these Norm-based Greedy heuristics use FFDAvgSum which are exp(x), which is the value of the exponential function at the point x, where x is 
average of sum of demand resources (e.g. CPU, memory, storage, network bandwidth, etc.).
VBP-Norm-LX assigns new VM to any host that has minimum of these norm values.

\item
EPOBF-ST and EPOBF-FT, presented in \cite{Quang-Hung2014}. The EPOBF-ST and EPOBF-FT algorithms sorts the list of $VM_{i}$ (i=1, 2,..., n) by their starting time ($ts_{i}$) and respectively by their finished time ($ts_{i}+dur_{i}$). Both EPOBF-ST and EPOBF-FT choose a host that has maximum of performance-per-watt to assign a new VM. The performance-per-watt is ratio of total of maximum capacity MIPS and maximum host's power consumption.

\item
MinDFT-ST and MinDFT-FT, presented in \cite{HungFDSE2014}. The MinDFT-ST and MinDFT-FT algorithms sorts the list of $VM_{i}$ (i=1, 2,..., n) by their starting time ($ts_{i}$) and respectively by their finished time ($ts_{i}+dur_{i}$). Both MinDFT-ST and MinDFT-FT allocate each VM (in a given set of VMs) to a host that has a minimum increase in total completion times of hosts.

\item
EMinRET, our proposed algorithm discussed in Section \ref{sec:schedalgo}. We evaluate the EMinRET with some its configurations:
 The EMinRET-1 sorts the list of VMs by VM's earliest starting time first and host's allocated VMs by its starting time.
 The EMinRET-2 sorts the list of VMs by VM's earliest starting time first and host's allocated VMs by its finishing times.
 The VM's finishing time, which is sum of its starting time and its duration time, is calculated by ($ts_{i}+dur_{i}$).
 The EMinRET-3 sorts the list of VMs by VM's earliest finishing time first and host's allocated VMs by its starting time.
 The EMinRET-4 sorts the list of VMs by VM's earliest finishing time first and host's allocated VMs by its finishing time.
 The EMinRET-5 sorts the list of VMs by VM's longest duration time first and host's allocated VMs by its starting time.
 The EMinRET-6 sorts the list of VMs by VM's longest duration time first and host's allocated VMs by its finishing time.
 The EMinRET-7 sorts the list of VMs by VM's latest finishing time first and host's allocated VMs by its starting time.
 The EMinRET-8 sorts the list of VMs by VM's latest finishing time first and host's allocated VMs by its finishing time.

\end{itemize}

\subsection{Methodology}

\input{vm-types}

\input{host-power}

\input{sim-result}

\begin{figure}[!ht]
    \centering
	\includegraphics[width=0.45\textwidth,height=4.4cm]{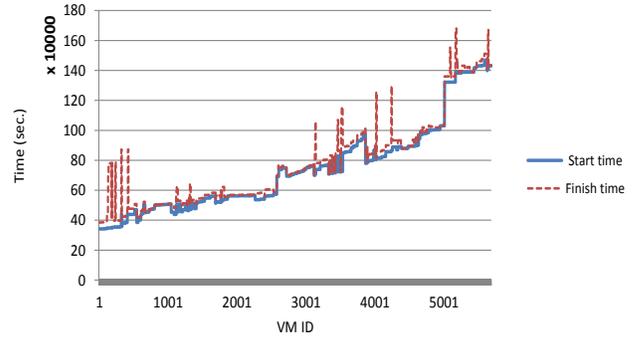}
    \caption[VM's starting and finishing time of VMs]{Starting time (blue line) and finishing time (dotted red line) of VMs in simulations with HPC2N Seth log-trace \cite{HPC2NWorkload}.}
    \label{fig:vmtime}
\end{figure}


\begin{figure}[!ht]
\centering
\includegraphics[width=0.45\textwidth,height=4.4cm]{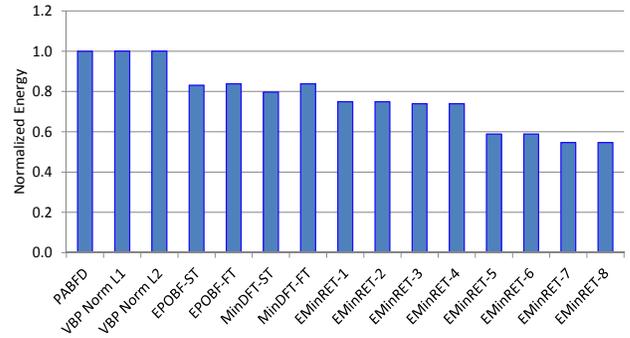}
\caption{Normalized energy. Result of simulations with HPC2N Seth log-trace.}
    \label{fig:scaledenergyresult-hpc2n}
\end{figure}

\begin{figure}[!ht]
    \centering
    \includegraphics[width=0.45\textwidth, height=4.4cm]{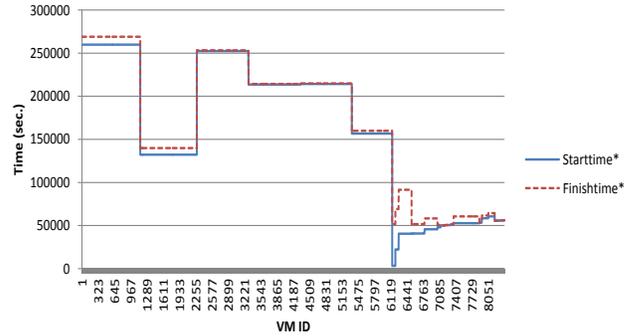}
    \caption[Starting time of VMs]{Starting time (blue line) and finishing time (dotted red line) of VMs in simulations with SDSC Blue Horizon log-trace \cite{SDSCBLUEWorkload}. Both starting time* and finishing time* in the chart which has their values from the simulated starting and finishing time subtracts 390,000 seconds.}
    \label{fig:vmtime-sdscblue}
\end{figure}



\begin{figure}[!ht]
\centering
\includegraphics[width=0.45\textwidth,height=4.4cm]{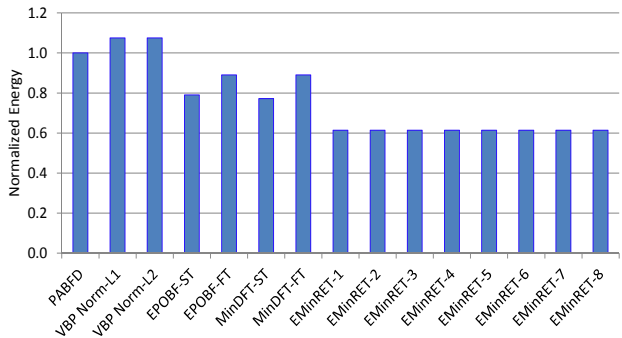}
\caption{Normalized energy. Result of simulations with SDSC Blue Horizon log-trace.}
    \label{fig:chart-scaledenergy-sdsc-blue}
\end{figure}

We evaluate these algorithms by simulation using the CloudSim \cite{Cloudsim} to create a simulated
 cloud data center system that has identical physical machines, heterogeneous VMs,
 and with thousands of CloudSim's cloudlets \cite{Cloudsim} (we assume that each
HPC job's task is modeled as a cloudlet that is run on a single VM). 
The information of VMs (and also cloudlets) in these simulated workloads 
is extracted from two real log-traces ( SDSC Blue Horizon log-trace \cite{SDSCBLUEWorkload}) and 
HPC2N Seth log-trace \cite{HPC2NWorkload}) 
in Feitelson's Parallel Workloads Archive (PWA) \cite{PWA} to 
model HPC jobs. When converting from the 
log-trace, each cloudlet's length is a product of the system's 
processing time and CPU rating (we set the CPU rating is equal to included VM's MIPS).
 We convert job's submission time, job's start time (if the start time is missing, then the start time is equal to sum of job's submission time and job's waiting time), 
 job's request run-time, 
 and job's number of processors 
 in job data from the log-trace in the PWA
 to VM's submission time, starting time and duration time, 
 and number of VMs (each VM is created in round-robin in the four types of VMs in Table \ref{tab:vmtype} on the number of VMs). 
 Four types of VMs as presented in the Table \ref{tab:vmtype} is similar to categories in Amazon EC2's 
 VM instances: high-CPU VM, high-memory VM, small VM, and micro VM. 
 Figure \ref{fig:vmtime} shows chart of starting times and finishing times
 of the VMs in a simulation (the simulations have the same starting times and duration times of VMs).
 All physical machines are identical and each physical machine has x4 CPU cores (2660 MIPS per core),
 8192 MB of physical memory, 10 Gb/s of network bandwidth, 1 TBytes of storage.
 Power model of each physical machine is shown in Table \ref{tab:HostPower},
 which is from a typical Hewlett-Packard Company ProLiant ML110 G5 in SPECpower\_ssj2008 benchmark \cite{HPProliantG5PowerModel}.
In the simulations, we use weights as following:
(i) weight of increasing time of mapping a VM to a host: \{0.001, 0.01, 1, 100, 3600\};
(ii) weights of computing resources such as number of MIPS per CPU core, physical memory (RAM), network bandwidth, and storage respectively: 940, 24414, 1, 0.0001 respectively. 
We will discuss how to choose these values for weights of resources in another paper. 
We simulate on combination of these weights. 
The total energy consumption of each EMinRET-[1-8] is the average of five times simulation with various weights of increasing time (e.g. 0.001, 0.01, 1, 100, or 3600).

We choose PABFD \cite{Beloglazov2011} as the baseline algorithm 
because the PABFD is a famous power-aware best-fit decreasing
in the energy-aware scheduling research community.
We also compare our proposed VM 
allocation algorithms with two vector bin-packing algorithms (VBP-Norm-L1/L2) to show 
the importance of with/without considering VM's starting time and finish time 
in reducing the total energy consumption of VM placement problem.


\subsection{Results and Discussions}
\label{sec:resultsanddiscussions}

The Table \ref{table:simresult-hpc2n} shows simulation results of scheduling algorithms solving scheduling problems 
 with 5,687 VMs and 10,000 physical machines (hosts),
 in which VM's data is converted from the HPC2N Seth log-trace \cite{HPC2NWorkload}.
 The Table \ref{table:simresult-sdscblue} shows simulation results of scheduling algorithms solving scheduling problems 
  with 8,368 VMs and 5,000 physical machines (hosts),
  in which VM's data is converted from the SDSC BLUE log-trace \cite{SDSCBLUEWorkload}.
Both Figure \ref{fig:scaledenergyresult-hpc2n} and Figure \ref{fig:chart-scaledenergy-sdsc-blue} show a bar chart comparing energy consumption of VM allocation algorithms that scale with the PABFD.
None of the algorithms use VM migration techniques,
and all of them  satisfy the Quality of Service
(e.g. the scheduling algorithm provisions maximum of user VM's requested resources).
We use total energy consumption as the performance metric for
evaluating these VM allocation algorithms.
The energy saving shown in both Table \ref{table:simresult-hpc2n} and Table \ref{table:simresult-sdscblue} are the reduction of
total energy consumption of the corresponding algorithm
compared with the baseline PABFD  \cite{Beloglazov2011} algorithm.

Table \ref{table:simresult-hpc2n} shows that, compared with PABFD \cite{Beloglazov2011},
 our EMinRET with 8 configurations
 can reduce the total energy consumption by average 34.25\%, and
 average 34.25\% 
 respectively compared with
 norm-based vector bin-packing algorithms
 (VBP-Norm-L1/L2) in simulations with the first 300 jobs of the HPC2N Seth log-trace. 
 Table \ref{table:simresult-sdscblue} shows that, compared with PABFD \cite{Beloglazov2011} 
 or norm-based vector bin-packing algorithms (e.g. VBP-Norm-L2) \cite{Panigrahy2011},
 our EMinRET-[1-8]
 reduces the total energy consumption by average 39\%
 in simulations with the first 50 jobs in the SDSC BLUE log-trace.


The PABFD generates a schedule that uses higher energy consumption than
the MinRET-ST and EMinRET-FT because of the following main reasons.
First, our hypothesis in this paper is that each VM consumes the same amount of energy 
in any physical server ($e_{i}$) and all physical servers
are identical. In consequence, the PABFD will choose a random
physical server to map a new VM. 
The PABFD sorts the list of VMs by decreasing requested computing power (e.g. MIPS), 
 therefore the PABFD allocates VMs that has the most requested computing power firstly. 
 In Table \ref{tab:vmtype}, a VM1-typed VM has highest requested computing power in the list, next is VM2-typed VMs, etc..
Instead, our proposed EMinRET-[1-8] algorithms assign a new VM to a physical 
server in such a way that has minimum increase of completion times and use fully all resources in physical machines.

These EMinRET-[1-8] algorithms perform
better than our previous algorithms such as MinDFT-ST/FT and EPOBF-ST/FT in the simulations. 
Compared to EPOBF-ST and EPOBF-FT, the EMinRET-[1-8] have less
total energy consumption than by average 14.2\% and respectively average 14.9\%.
The EMinRET-[1-8] have also less
total energy consumption than the MinDFT-ST and MinDFT-FT from average 10.6\% and respectively 14.9\%.
In the simulations, swapping between a new VM and its overlapped VM that is allocated to a host 
 reduce total completion time on the host. 
 For input as in Table \ref{table:example1}, the VM2 is remove from the first host, 
 the VM6 will allocated to the first host.

%% file: vm-types.tex
\begin{table}[htp]
\caption{Four VM type in simulations}
\label{tab:vmtype}
\centering

\begin{tabular}{|l|c|c|c|c|c|}
\hline
Type                    & \multicolumn{1}{l|}{Cores} & \multicolumn{1}{l|}{MIPS} & \multicolumn{1}{l|}{Mem. (MB)} & \multicolumn{1}{l|}{Net. (Mb/s)} & \multicolumn{1}{l|}{Storage (GB)} \\ \hline
VM1                     & 2                              & 2500                     & 871                             & 100                             & 5                             \\
VM2	 					& 1                              & 2000                     & 3840                            & 100                             & 5                             \\
VM3                     & 1                              & 1000                     & 1536                            & 100                             & 5                             \\
VM4                     & 1                              & 500                       & 613                             & 100                             & 5                             \\ \hline
\end{tabular}
\end{table}

%% file: host-power.tex
\begin{table*}[htp]
\caption{Host power consumption model with CPU utilization of an typical server with 4 cores, 8192 MBytes of RAM, 10Gb/s of network bandwidth, 1TBytes of storage.}
\label{tab:HostPower}
\centering
\begin{tabular}{|l|r|r|r|r|r|r|r|r|r|r|r|}
\hline
CPU Utilization (\%) & 0\%  & 10\% & 20\% & 30\% & 40\% & 50\% & 60\% & 70\% & 80\% & 90\%  & 100\% \\ \hline
Host power (Watts)   &  93.7 &  97.0 & 101.0 & 105.0 & 110.0 & 116.0 & 121.0 & 125.0 & 129.0 & 133.0 & 135.0 \\ \hline
\end{tabular}

\end{table*}

%% file: sim-result.tex
\begin{table}[!h]
\centering
\caption{Result of simulations using the first 300 jobs of the HPC2N Seth log-trace \cite{HPC2NWorkload}}
\label{table:simresult-hpc2n}
\resizebox{0.5\textwidth}{!}{
\begin{tabular}{|l|c|c|c|c|}
\hline
Algorithm & \multicolumn{1}{l|}{\#Hosts} & \multicolumn{1}{l|}{\#VMs} & \multicolumn{1}{l|}{Energy (KWh)} & \multicolumn{1}{l|}{Saving (\%)} \\ \hline
(1) PABFD \cite{Beloglazov2011} (baseline)  & 10000 & 5687 & 3325.72 & 0\% \\ \hline
(2) VBP Norm L1 \cite{Panigrahy2011} & 10000 & 5687 & 3328.81 & 0\% \\ \hline
(3) VBP Norm L2 \cite{Panigrahy2011} & 10000 & 5687 & 3328.81 & 0\% \\ \hline
(4) EPOBF-ST \cite{Quang-Hung2014} & 10000 & 5687 & 2763.78 & 17\% \\ \hline
(6) EPOBF-FT \cite{Quang-Hung2014} & 10000 & 5687 & 2786.90 & 16\% \\ \hline
(7) MinDFT-ST \cite{HungFDSE2014} & 10000 & 5687 & 2651.58 & 20\% \\ \hline
(8) MinDFT-FT \cite{HungFDSE2014} & 10000 & 5687 & 2786.90 & 16\% \\ \hline
(9) EMinRET-1 & 10000 & 5687 & 2491.70 & 25\% \\ \hline
(10) EMinRET-2 & 10000 & 5687 & 2490.81 & 25\% \\ \hline
(11) EMinRET-3 & 10000 & 5687 & 2457.73 & 26\% \\ \hline
(12) EMinRET-4 & 10000 & 5687 & 2457.73 & 26\% \\ \hline
(13) EMinRET-5 & 10000 & 5687 & 1957.67 & 41\% \\ \hline
(14) EMinRET-6 & 10000 & 5687 & 1957.99 & 41\%  \\ \hline
(15) EMinRET-7 & 10000 & 5687 & 1817.62 & 45\%  \\ \hline
(16) EMinRET-8 & 10000 & 5687 & 1817.62 & 45\%  \\ \hline
\end{tabular}
}
\end{table}

\begin{table}[!h]
\centering
\caption{Result of simulations using the first 50 jobs of SDSC BLUE log-trace \cite{SDSCBLUEWorkload}.}
\label{table:simresult-sdscblue}
\resizebox{0.5\textwidth}{!}{ 
\begin{tabular}{|l|c|c|c|c|}
\hline
Algorithm & \multicolumn{1}{l|}{\#Hosts} & \multicolumn{1}{l|}{\#VMs} & \multicolumn{1}{l|}{Energy (KWh)} & \multicolumn{1}{l|}{Saving (\%)} \\ \hline
(1) PABFD  \cite{Beloglazov2011} (baseline)  & 5000 & 8368 & 729.51 & 0\% \\ \hline
(2) VBP Norm-L1 \cite{Panigrahy2011} & 5000 & 8368 & 784.04 & -7\% \\ \hline
(3) VBP Norm-L2 \cite{Panigrahy2011} & 5000 & 8368 & 784.04 & -7\% \\ \hline
(4) EPOBF-ST \cite{Quang-Hung2014} & 5000 & 8368 & 576.59 & 21\% \\ \hline
(5) EPOBF-FT \cite{Quang-Hung2014} & 5000 & 8368 & 649.23 & 11\% \\ \hline
(6) MinDFT-ST \cite{HungFDSE2014} & 5000 & 8368 & 563.27 & 23\% \\ \hline
(7) MinDFT-FT \cite{HungFDSE2014} & 5000 & 8368 & 649.23 & 11\% \\ \hline
(8) EMinRET-1 & 5000 & 8368 & 447.84 & 39\% \\ \hline
(9) EMinRET-2 & 5000 & 8368 & 447.84 & 39\% \\ \hline
(10) EMinRET-3 & 5000 & 8368 & 447.84 & 39\% \\ \hline
(11) EMinRET-4 & 5000 & 8368 & 447.84 & 39\% \\ \hline
(12) EMinRET-5 & 5000 & 8368 & 447.84 & 39\% \\ \hline
(13) EMinRET-6 & 5000 & 8368 & 447.84 & 39\% \\ \hline
(14) EMinRET-7 & 5000 & 8368 & 447.84 & 39\% \\ \hline
(15) EMinRET-8 & 5000 & 8368 & 447.84 & 39\% \\ \hline
\end{tabular}
}
\end{table}

%% file: concl.tex
In this paper, we formulated an energy-aware VM allocation problem with fixed starting time and non-preemption.
We also discussed our two key observations in the VM allocation problem.
First, minimizing total energy consumption is equivalent to
minimizing the sum of total completion time of all physical machines (PMs).
For some possible schedules, which have same total completion time of all PMs, 
the RET metric decides a schedule that has higher resource efficiency.
Second, swapping between an unallocated VM and its overlapped VM, which has already been allocated to a PM, can 
 reduce the total completion time of all PMs.
Based on these observations,
we proposed EMinRET algorithm to solve the energy-aware VM allocation with fixed starting time and duration time.
Our proposed EMinRET and its sorting list of VMs by starting time (or longest duration time first, or latest finishing time first)
can all reduce the total energy consumption
of the physical servers 
compared with other algorithms in 
simulation results on the HPC2N Seth \cite{HPC2NWorkload} 
and SDSC Blue Horizon log-traces \cite{SDSCBLUEWorkload} in the Feitelson's PWA \cite{PWA}. 
The combination of EMinRET with its sorting list of VMs by latest finishing time first (EMinRET-8) is the best.


In future, we are developing EMinRET into a cloud resource management software (e.g. OpenStack Nova Scheduler).
Additionally, we are working on IaaS cloud systems with heterogeneous physical servers and job requests consisting of multiple VMs.
We are studying how to choose the right weights of time and resources 
(e.g. computing power, physical memory, network bandwidth, etc.) in Machine Learning techniques.
